\documentclass{sig-alternate}

\conferenceinfo{HT'11,} {June 6--9, 2011, Eindhoven, The Netherlands.}
\CopyrightYear{2011}
\crdata{978-1-4503-0256-2/11/06}
\usepackage{subfigure}
\usepackage{url}
\usepackage{multirow}

\newenvironment{packed_description}{
\begin{description}
 \setlength{\itemsep}{1pt}
 \setlength{\parskip}{0pt}
 \setlength{\parsep}{0pt}
}{\end{description}}

\begin{document}
\title{Modeling the Structure and Evolution of Discussion Cascades}
%\title{Alternate {\ttlit ACM} SIG Proceedings Paper in LaTeX
%Format\titlenote{(Produces the WWW2010-specific release, location and
%copyright information). For use with
%www2010-submission.cls V1.4. Supported by ACM.}}
%\subtitle{[Extended Abstract]
%\titlenote{A full version of this paper is available as
%\textit{Author's Guide to Preparing ACM SIG Proceedings Using
%\LaTeX$2_\epsilon$\ and BibTeX} at
% You need the command \numberofauthors to handle the "boxing"
% and alignment of the authors under the title, and to add
% a section for authors number 4 through n.
%
% Up to the first three authors are aligned under the title;
% use the \alignauthor commands below to handle those names
% and affiliations. Add names, affiliations, addresses for
% additional authors as the argument to \additionalauthors;
% these will be set for you without further effort on your
% part as the last section in the body of your article BEFORE
% References or any Appendices.

\numberofauthors{3}
%
% Put no more than the first THREE authors in the \author command

% NOTE: All authors should be on the first page. For instructions
% for more than 3 authors, see:
% http://www.acm.org/sigs/pubs/proceed/sigfaq.htm#a18

\author{
%% e-mail address with \email.
\alignauthor Vicen\c{c} G\'omez\\
       \affaddr{Donders Institute for Brain Cognition and Behaviour}\\
       \affaddr{Radboud University Nijmegen}\\
       \email{\normalsize{v.gomez@science.ru.nl}}
\alignauthor Hilbert J. Kappen\\
       \affaddr{Donders Institute for Brain Cognition and Behaviour}\\
       \affaddr{Radboud University Nijmegen}\\
       \email{\normalsize{bertk@science.ru.nl}}
\alignauthor Andreas Kaltenbrunner\\
       \affaddr{Information, Technology and Society Research Group}\\
       \affaddr{Barcelona Media}\\
       \email{\normalsize{kaltenbrunner@gmail.com}}
}
%\additionalauthors{Additional authors: John Smith (The Th{\o}rv\"{a}ld Group,
%email: {\texttt{jsmith@affiliation.org}}) and Julius P.~Kumquat
%(The Kumquat Consortium, email: {\texttt{jpkumquat@consortium.net}}).}
\date{30 July 1999}

\newcommand{\fix}{\marginpar{FIX}}
\newcommand{\new}{\marginpar{NEW}}
\newcommand{\bc}{\begin{center}}
\newcommand{\ec}{\end{center}}
\newcommand{\bd}{\begin{description}}
\newcommand{\ed}{\end{description}}
\newcommand{\zp}{\ensuremath{(e^\beta d_1)^{\alpha_1}}}
\newcommand{\zpl}{\ensuremath{(\beta + \log d_1)}}
\newcommand{\zc}{\ensuremath{\sum_{l=2}^t{d_l^{\alpha_c}}}}  
\newcommand{\zcp}{\ensuremath{\left(\sum_{l=2}^t{d_l^{\alpha_c}}\right)}}  
\newcommand{\zcl}{\ensuremath{\sum_{l=2}^t{d_l^{\alpha}\log d_l}}}  
\newcommand{\zclp}{\ensuremath{\left(\sum_{l=2}^t{d_l^{\alpha}\log d_l}\right)}}  
\newcommand{\zcll}{\ensuremath{\sum_{l=2}^t{d_l^{\alpha}\log^2 d_l}}}  

\newcommand{\ak}{\ensuremath{{\alpha_k}}}
\newcommand{\bk}{\ensuremath{{\beta_k}}}
\newcommand{\al}{\ensuremath{{\alpha_l}}}
\newcommand{\bl}{\ensuremath{{\beta_l}}}

\maketitle
\begin{abstract}
  % AK better don't use emph in the abstract. Barrapunto and Meneame
  % may not be well know, maybe we should describe them
We analyze the structure and evolution of discussion cascades in four popular
websites: \emph{Slashdot}, \emph{Barrapunto}, \emph{Meneame} and
\emph{Wikipedia}. Despite the big heterogeneities between these sites, a
preferential attachment (PA) model with bias to the root can capture the
% AKht maybe it is no simple
temporal evolution of the observed trees and many of their statistical
properties, namely, probability distributions of the branching factors
(degrees), subtree sizes and certain correlations. The
parameters of the model are learned efficiently using a novel maximum
% AKht is it novel? Ref 23 is similar?
% Vicen: they use another model...
likelihood estimation scheme for PA and provide a figurative interpretation
about the communication habits and the resulting discussion cascades on the
four different websites.
\end{abstract}

% A category with only the three required fields
%\category{H.4.m}{Information Systems}{Miscellaneous}\\
%
%\category{H.4.m}{Information Systems}{Miscellaneous}
%\category{D.2}{Software}{Software Engineering}
%%A category including the fourth, optional field follows...
%\category{D.2.8}{Software Engineering}{Metrics}[complexity measures,
%performance measures]



\section{Introduction}
Human communication patterns on the Internet are characterized by transient
responses to social events.  Examples of such phenomena are the discussion
threads generated in news aggregators, the propagation of massively circulated
Internet chain letters, or the synthesis of articles in collaborative web-based
spaces such as Wikipedia.

\begin{figure*}[!th]
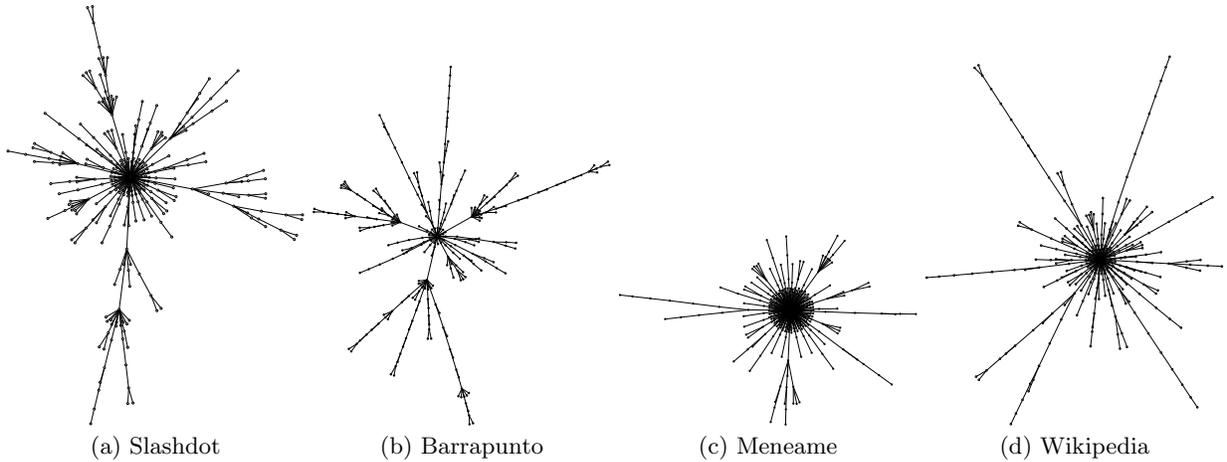

\begin{center}
  \subfigure[Slashdot]{\includegraphics[width=.47\columnwidth]{figures/sl_tree.dot.eps}}
  \subfigure[Barrapunto]{\includegraphics[width=.47\columnwidth]{figures/bp_tree.dot.eps}}
  \subfigure[Meneame]{\includegraphics[width=.47\columnwidth]{figures/mn_tree.dot.eps}}
  \subfigure[Wikipedia]{\includegraphics[width=.47\columnwidth]{figures/wk_tree.dot.eps}}
\end{center}
\caption[Example]{Examples of real discussion cascades.
%[Describe them? Title, num. comments, num. users, etc]
}
\label{fig:real}
\end{figure*}

These responses can be regarded as tree-like cascades of activity
generated from an underlying social network.  Typically, a trigger
event, or a small set of initiators, generate a chain reaction which
may catch the attention of other users who end up participating in the
cascade (see Figure \ref{fig:real} for examples) and attract even more
users.  Since these cascades of comments are a direct consequence of
the information flow in a social system, understanding the mechanisms
and patterns which govern them plays a fundamental role in contexts
like spreading of technological innovations \cite{rogers}, diffusion
of news and opinion \cite{gruhl,Leskovek07}, social influence
\cite{Bakshy} %viral marketing \cite{viral}
or collective problem-solving \cite{problems}.

Although information
cascades have been extensively analyzed for particular domains, such
as blogs \cite{gruhl,Leskovek07}, chain letters \cite{Nowell08},
Flickr \cite{flickr}, Twitter \cite{twitter} or page diffusion on
Facebook \cite{facebook}, the cascades under consideration in those
studies rarely involve elaborated discussions or complex interchange
of opinions: generally, a small piece of information is just forwarded
from an individual to its direct neighbors.  To the best of our
knowledge, with the exception of \cite{Kumar}, no previous work exists
on modeling the evolution and structure of long discussion-based
cascades.

%It remains as a open question whether the spread of information in
%discussion-based cascades follows similar patterns and is governed by the same
%mechanisms.
Here, as in \cite{Kumar}, we consider several websites where
the associated (discussion) cascades contain high level of interaction.
We analyze for the first time the cascades of the popular news aggregator 
%AKht added of
\emph{Slashdot}, \emph{Barrapunto} (a Spanish version of Slashdot) and
\emph{Meneame} (a Spanish \emph{Digg}-clone) and the English \emph{Wikipedia}.
As the reader may notice, these datasets are quite heterogeneous.  For
%I would not speak about the datasets, rather about the architecture of the corresponding websites
instance, although posts from both Slashdot and Meneame correspond to popular
news which rely on broadcasted events, Slashdot contains rich and very
extensive comments, which are less frequent in Meneame.  
%AKht This comment has no fundament, better speak about the architecture
% Vicen: ok, todo
The cascades found in
Wikipedia, on the other hand, represent collaborative effort towards a well
defined goal: produce a free, reliable article.

In this study we address the following questions: what are the
statistical patterns that determine the structure of such cascades and
their evolution?  Can these patterns be largely determined regardless
of semantic information using a simple parametric model? Can the
parameterization corresponding to a given website provide a global
characterization for it?

% AK: None follows a scale free distribution.
We first provide a global analysis of the cascade behavior in the four
mentioned websites. Among other results, we find that typically, the sizes of
the cascades have a clear defined scale, which seems to contradict the recent
results of \cite{Kumar}.  Our analysis also highlights the importance of
repetitive user participation in relation to other types of cascades and their
impact on the entire social network.  We also present a growth model for
discussion cascades which is validated in the four datasets.  Our approach is
based on a simple model of preferential attachment (PA) \cite{barabasi99a},
where new contributions in the cascade tree are linked to existing
contributions with a probability which depends on their popularity (degree). 
%AK better this way to avoid problems with the negative alphas for WP.
%are preferentially linked to existing contributions with high popularity
%(degree).  

Two key ingredients characterize our approach: First, we account for a certain
bias favoring the root, or event initiator.  In this way, we are able to
capture the different processes governing the global (direct reactions) and the
localized responses of the system.  Second, we use a likelihood method
particularly developed for this study which allows an efficient estimation of
the model parameters which considers the \emph{entire} generative model.  The
method is applicable not only for the data considered here but for a more
general class of growing graphs.  
% AK only trees?
%It is important to emphasize that we do not explicitly model the
%network dynamics over which the diffusion happens (dynamics at the
%level of the user). We focus only on the stochastic process which
%generates the cascade.
% AKht Better:
 Here we are only interested on the stochastic process which generates
 the cascade. We do not model network dynamics or a termination
 criteria for the cascades.  Such a model could be built on top of our
 current model as it is done for example in \cite{Kumar}.

%This paper is organized in the following way.
In the next Section, we explain the proposed model and how we estimate its
parameters.  Section \ref{sec:datasets} introduces the datasets and provides a
global analysis about their main characteristics. In Section~\ref{sec:results}
we explain the main results and give an interpretation of the parameters of the
model.  Finally, in Section~\ref{sec:related} we describe related work and
discuss the results in Section~\ref{sec:discussion}.  In the Appendix we
explain some aspects of the likelihood approach which are important for the
estimation of parameters.

% AK, Vicen 
%\end{itemize}
%\begin{itemize}
\section{Growing tree models for discussion cascades}
\label{sec:model}
We model a discussion cascade as a growing network in which nodes correspond to
comments and the initial node corresponds to the post (a news article, for
instance).  A new node is added sequentially at discrete time-steps.  Our model
is based on the original PA model to which we add a bias to the first node.
Since each new node adds only one new link to the existing graph, the resulting
network is a tree.  We also assume that the total number of nodes $N$ is known.
It is convenient to represent compactly the cascade as a vector of parent nodes
$\boldsymbol{\pi}$, where $\pi_t$ denotes the parent of the node $t+1$ added at
time-step $t$. %AKfw at time t+1 or at time t?
%better: ... the parent of the node $t+1$ added at time-step $t$.

%associated tree structure.  In a PA model, new nodes are
%preferentially linked to existing contributions with high popularity (degree).
%each new node is attached to a
%target node selected with probability proportional to its degree. The process
%is repeated until $N$ nodes are added to the network. 
%where the root of the tree corresponds
%to the initiating event, for
%instance, the news post i.  

We are interested in the probability of being node $k$ the parent
$\pi_{t}$ given the past history $\boldsymbol{\pi}_{(1:t-1)}$,
that is $p(\pi_t = k| \boldsymbol{\pi}_{(1:t-1)})$, for
$t>1$, $k=\{1,\hdots,t\}$ %AKfw I think it is $k=[1,\hdots,t]$
and initial vector $\boldsymbol{\pi}_1=(1)$
\footnote{ At time $0$ we have $\boldsymbol{\pi}_0=()$ and for all
  trees, $p(\pi_1=1) = 1$ and $0$ otherwise, i.e.
  $\boldsymbol{\pi}_1=(1)$ always.}. Note that by construction, $\pi_t
\leq t, \forall t$.

At time-step $t$, we relate the \emph{popularity} of a node $k$ with its number
of links (degree $d_{k,t}$) before node $t+1$ is added in the following way:
\begin{align}
d_{k,t} ( \boldsymbol{\pi}_{(1:t-1)})&=
\begin{cases}
1+\sum_{m=2}^{t-1}{\delta_{k\pi_m}} & \text{for $k\in\{1,\hdots,t\}$}\\
0 & \text{otherwise}
\end{cases},
\label{eq:degree}
\end{align}
%$_{k,t}(\boldsymbol{\pi}_{(1:t)})$ and is the number of times $k$
%occurs in $\boldsymbol{\pi}_{(1:t)}$.
%\end{align}
where $\delta$ is the Kronecker delta function.
%and $c$ is usually denoted as the \emph{initial attractiveness} of a node. 
In the following, we omit the explicit dependence on $\boldsymbol{\pi}_{(1:t-1)}$, so that
%we assume $c=1$ and 
$d_{k,t}\equiv d_{k,t}( \boldsymbol{\pi}_{(1:t-1)})$.
\begin{figure}[!h]
\begin{center}
 \includegraphics[width=.7\columnwidth]{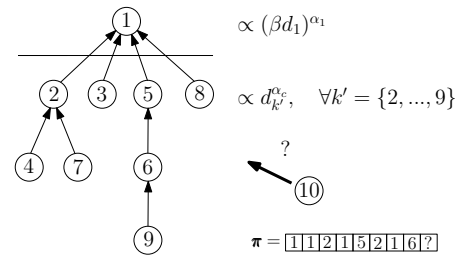}
\end{center}
\caption[Example] {Small example: at time-step $9$, node number $10$
  is added to the cascade.  With probability proportional to $(\beta
  d_1)^{\alpha_1}$ it is added to the root node (initiator) and with
  probability proportional to $d_{k'}^{\alpha_c}$ to one of the 
  non-root nodes.  Bottom right shows the corresponding parent vector
  $\boldsymbol{\pi}$ (see text for definitions).  }
\label{fig:model}
\end{figure}

The PA model attaches new nodes to node $k$ with probability proportional to
its popularity.  See Figure \ref{fig:model} for an illustration.  For
completeness, we consider two models: a simple PA model \emph{without bias} to
the root and a model which differentiates between the root node and the rest.

In the general PA model, the probability for attaching the node $t+1$ to node
$k$ at time-step $t>1$ is parameterized using a linear term $\beta_k$ and an
exponent $\alpha_k$ for each of the nodes:

%At time $t+1$, the PA model attaches the new node to a node $k$ with
%probability proportional to its popularity.  See Figure \ref{fig:model} for an
%illustration.  For completeness, we consider three models: a simple PA %AKht changed
%model \emph{without bias} to the root, a model which differentiates between
%the root node and the rest, and the T-MODEL introduced in \cite{Kumar}. In the
%general PA model the probability for attaching the node $t+1$ to node $k$ can be
%written as:
%(using $Z_t$ is the normalization factor):
%probability for attaching the node $t+1$ to the root is is
%proportion$\propto(\exp(\beta)d_{1,t})^{\alpha_1}$ and the probability for attaching it to a
%non-root node $c$ is $\propto(d_{c,t})^{\alpha_c}$. 
\begin{align}
\label{eq:model}
p(\pi_{t} = k| \boldsymbol{\pi}_{(1:t-1)})
&= 
\frac{1}{Z_t}(\beta_{k}d_{k,t})^{\alpha_k},&
Z_t & = \sum_{l=1}^{t}{ (\beta_l d_{l,t}})^{\alpha_l}.
\end{align}
%\begin{description}
\textbf{Model without bias}: 
If we set $\alpha_k=\alpha$ and $\beta_k=1$, for $k=\{1,\hdots,t\}$, we recover
an important generalization of Barabasi's PA model, where the probability of
attachment to a node goes as some general power $\alpha$ of the degree
\cite{kaprivsky, jeong}.  For $\alpha = 1$, the linear preferential attachment
is recovered. In this case, nodes have power-law distributed degrees.  For
$\alpha~<~1$, or sublinear PA, the degrees are distributed according to a
stretched exponential.  For $\alpha > 1$ there is a ``condensation'' phenomenon,
in which a single node gets a finite fraction of all the connections in the
network \cite{kaprivsky}.

\noindent\textbf{Model with bias}: 
Consider the following parameterization:
\begin{align}
\label{eq:parameterization}
\small
\alpha_k&=
\begin{cases}
\alpha_1 & \text{for $k=1$}\\
\alpha_c & \text{for $k\in\{2,\hdots,t\}$}
\end{cases}\notag\\
\beta_k&=
\begin{cases}
\beta & \text{for $k=1$}\\
1     & \text{for $k\in\{2,\hdots,t\} $}
\end{cases}.
\end{align}
In this case, $\alpha_1$ and $\alpha_c$ are the exponents of the PA
processes governing the root and the non-root nodes respectively.
$\beta$ can be regarded as an additional degree of freedom weighting the root
of the tree. In Section \ref{sec:params} we discuss about the interpretability of
these parameters. 
%In this work, we use a simplified model which considers two parameters for the
%root: $\beta_1$ and $\alpha_1$ and only one parameter for the rest of the nodes
%$\alpha_k = \alpha_c$ ($\beta_c = 1$), for $c>1$.  Thus, the probability of
%attaching the new node to the root is $p(\pi_{t+1}=1|\boldsymbol{\pi}_{(1:t)})
%\propto \beta_pd_1^{\alpha_p}$ and $p(\pi_{t+1} \in |\boldsymbol{\pi}_{(1:t)} )
%\propto d_c^{\alpha_c}$ otherwise.  This allows to retain a minimal set of
%parameters while capturing the \emph{bias} towards the originator of the
%cascade, a fact that we often observed empirically.

Note that, although we explicitly model the event which triggers the cascade as
a root node, this representation does not limit the cascade to be originated
from an individual event only. The root node can of course represent a group of
initiators.

%After development of this manuscript we learned of 
%which overlaps some of our results for other three datasets.
%\noindent\textbf{T-MODEL}: 
%The previous model shares many similarities with the recently proposed in
%\cite{Kumar}.  The main difference is that, instead of the probability model of
%Equation \eqref{eq:model}, their growth cascade model uses other three
%parameters ($\Gamma,\tau$ and $\delta$) and evolves according to:
%%AKht Better use the same parameter where possible e.g. beta instead of alpha
%% Vicen: but beta is the bias to the post, not the PA parameter in our case...
%%       itś a bit confusing...
%\begin{align}
%\label{eq:tmodel}
%p(\pi_{t+1} = k| \boldsymbol{\pi}_{(1:t)})
%&= \frac{1}{Z_t}\left(\Gamma d_{k,t}+\tau^{t+1-k}\right),\notag\\
%Z_t &
%    = \delta + \Gamma(t-1)+\frac{\tau(\tau^t-1)}{(\tau-1)}.
%\end{align}
%% = \delta+\sum_{l=1}^{t}{\alpha d_{l,t}+\tau^{t+1-l}}
%or finishes with probability $\delta/Z_t$, Parameter $\Gamma$ (named $\alpha$
%in their paper) captures the popularity of a node, parameter $\tau$ the novelty
%and $\delta$ the size of the cascade. The attractiveness of a comment is thus
%expressed as a combination of popularity and novelty. 
\subsection{Maximum likelihood parameter estimation}
\label{sec:ml}
%After introducing the model, we present our approach to estimate the optimal
%parameters given a set of data.
Usually, PA in evolving networks is measured by calculating the rate at which
groups of nodes with identical connectivity form new links during a small time
interval $\Delta t$ \cite{jeong, blasio}.  However, this approach is suitable
only for networks with many nodes
%measuring probabilities as a function of the degrees is complicated
%since both the size of the graph and the degrees themselves are changing over
%time.  Indeed, 
that are stationary in the sense that the number of nodes remain constant during the
interval $\Delta t$. This is not a reasonable assumption in our data, which is
often produced by a transient, highly nonstationary, response.

Another approach for parameter estimation relies on fitting a measured
property, for instance the degree distribution, for which an
analytical form can be derived in the model under consideration.  For
the PA model, extensive results exist with emphasis precisely on the
degree distributions \cite{bennaim}.  However, two important aspects
are worth to mention here.  First, analytical results usually rely on
assumptions like a %AK removed the
continuum limit or on an infinite size network, which is also %AK added also
not the case
of our data.  Second, it is important to stress here that when
parameters are learned for a particular observable for which an
analytical form has been derived, the model may \emph{overfit} on this
measure, introducing a bias in other structural properties such as
subtree sizes, average depths, or other correlations.

Our approach considers the likelihood function corresponding to the
\emph{entire} generative process (instead of particular measures such as degree
distributions or subtree sizes) introduced before. We can assign to each
observation (each node arrival in each cascade) a given probability using
Equation \eqref{eq:model}. The parameters for which the likelihood is maximal
are the ones that best explain the data given the model assumptions (see
\cite{lik} for a similar approach for another network growth model).

Formally, we observe a set $\Pi:=\{\boldsymbol{\pi}_1, \hdots
\boldsymbol{\pi}_N\}$ of $N$ trees with respective sizes
$|\boldsymbol{\pi}_i|$, $i\in\{1,\hdots N\}$ and we want to find the values of
$\boldsymbol\theta:=\left(\alpha_1, \alpha_c, \beta\right)$ which best explain
the data. 
The likelihood function can be written as:
%\cite{jeong,Yule1925, Simon1955}
\begin{align}
\label{eq:lik}
\mathcal{L}(\boldsymbol{\Pi}| \boldsymbol\theta) 
&=\prod_{i=1}^{N}{p(\boldsymbol{\pi}_i| \boldsymbol\theta) }\notag\\
&=\prod_{i=1}^{N}\prod_{t=2}^{|\boldsymbol{\pi}_i|}{p(\pi_{t,i}|\boldsymbol{\pi}_{(1:t-1),i},\boldsymbol\theta) }\notag\\
&=\prod_{i=1}^{N}\prod_{t=2}^{|\boldsymbol{\pi}_i|}
(\beta_{x}d_{x,t,i})^{\alpha_x}
\left(\sum_{l=1}^{t} (\beta_l d_{l,t,i})^{\alpha_l}\right)^{-1},
%&=\prod_{i=1}^{N}\prod_{t=1}^{|\boldsymbol{\pi}_i|-1}\sum_{k=1}^{t+1}
%{\delta_{k,\pi_{t+1,i}}p(\pi_{t+1,i}=k|\boldsymbol{\pi}_{(1:t),i},\boldsymbol\theta) },
\end{align}
where $\boldsymbol{\pi}_{(1:t-1),i}$ is the vector of parents in the tree $i$
after time $t-1$, $x:=\pi_{t,i}$ is the parent of node $t+1$ in the tree $i$,
and $d_{x,t,i}:= d_{x,t}(\boldsymbol{\pi}_{(1:t-1),i})$ denotes the degree of
node $x$ as in Equation \eqref{eq:degree} in the tree $i$.
%where we use a simplified notation where $x\equiv\pi_{t+1,i}$ denotes the
%parent of comment $t+1$ in the tree $i$ and $\boldsymbol{\pi}_{(1:t),i}$ is the
%vector of parents in the tree $i$ until time $t$. 
Instead of maximizing \eqref{eq:lik} directly, it is more convenient to
minimize the negative of the log-likelihood function:
\begin{align}
\label{eq:loglikZ}
%-\log  \mathcal{L}(\boldsymbol{\Pi} | \boldsymbol\theta)  &=
\log \mathcal{L}(\boldsymbol{\Pi}| \boldsymbol\theta) & =
\sum_{i=1}^{N} \sum_{t=2}^{|\boldsymbol{\pi}_i|} 
\alpha_x(\log\beta_x+\log d_{x,t,i})-\log Z_{t,i}(\boldsymbol{\pi}_i|\boldsymbol{\theta}),
\end{align}
where $Z_{t,i}(\boldsymbol{\pi}_i|\boldsymbol{\theta})=\sum_{l=1}^{t}{ (\beta_l d_{l,t,i})^{\alpha_l}}$.

 For more details about the optimization see the Appendix.

%       \begin{align}\label{eq:loglik}
%       \log 
%       \mathcal{L}(\boldsymbol{\Pi} | \boldsymbol\theta) 
%       %&= \sum_{i=1}^{N} \sum_{t=2}^{|\boldsymbol{\pi}_i|-1}{\log{p(\pi_{i,t} = k| \boldsymbol{\pi}_{i,t-1}, \boldsymbol{\theta})}}\\
%       &= \sum_{i=1}^{N} \sum_{t=1}^{|\boldsymbol{\pi}_i|-1} 
%       %\sum_{k=1}^{t+1}\delta_{k,\pi_{t+1,i}}\left(
%       \alpha_k(\log\beta_k+\log d_{k,t,i})-
%       \log\displaystyle\sum_{l=1}^{t}{ (\beta_l d_{l,t,i})^{\alpha_l}}
%       \end{align}
%	 \log Z_{t,i}(\boldsymbol{\pi}_i,\boldsymbol{\theta}),\notag\\
%$d_{k,t,i}$ denotes the degree of node $k$
%in tree $i$ at time $t$. %, $\alpha_k = \alpha_1$ for $k=1$ and $\alpha_c$ otherwise
%and the corresponding partition sum $\log Z_{t,i}(\boldsymbol{\pi}_i,\boldsymbol{\theta})$ is:
%%of tree $i$ at time $t$ is:
%\begin{align}
%	Z_{i,t}(\boldsymbol{\pi}_i,\boldsymbol{\theta}) &=
%    \displaystyle\sum_{l=1}^{t}{ \beta_l d^{\alpha_l}_{l,i}}.
%\end{align}

\begin{table*}[!t] \centering
  \caption{Dataset statistics for Slashdot (SL), Barrapunto (BP), Meneame (MN) and Wikipedia (WK). }
\begin{tabular}{|c|r@{,}l|r@{,}l|r@{,}l|c|r@{,}l|c|}
\hline 
%    $\scriptsize{\text{dataset}}$ &
%    \multicolumn{2}{|c|}{$\scriptsize{\text{\#cascades}}$} &
%    \multicolumn{2}{|c|}{$\scriptsize{\text{\#nodes}}$} &
%    \multicolumn{2}{|c|}{$\scriptsize{\text{max. size}}$} &
%    $\scriptsize{\text{max. users}}$ & 
%    \multicolumn{2}{|c|}{$\scriptsize{\text{total users}}$} &
%    $\scriptsize{\text{rep. user}}$
    dataset &
    \multicolumn{2}{|c|}{\#cascades} &
    \multicolumn{2}{|c|}{\#nodes (comments)} &
    \multicolumn{2}{|c|}{max. nodes } &
    max. users & 
    \multicolumn{2}{|c|}{total users} &
    repeated user
\\\hline
SL    &   $9$ & $820$         &   $2,028$ & $518$    &    $1$ & $567$   &   $1,031$  &  $93$ & $638$     & $>1 \quad 99\%$  \\\hline
%BP    &  $7$ & $485$          &  $397$ & $148$       &    $1$ & $040$   &   $180$    &  $6$  & $864$     & $>1 \quad 85\%$  \\\hline
BP    &  $7$ & $485$          &  $357$ & $951$       &    \multicolumn{2}{|c|}{$841\quad$}    &   $180$    &  $6$  & $864$     & $>1 \quad 85\%$  \\\hline
MN    &  $58$ & $613$         &  $2,220$ & $714$     &    $2$ & $718$   &   $1,021$  &  $53$ & $877$     & 
$\begin{array}{cc}
    >1 & 34\% \\
    >5 & 70\%
 \end{array}$\\\hline 
%WK    &  $871$ & $485$        &  $9,421$ & $976$     &    $28$ & $701$  &  $3,616$  & $266$ & $493$    & 
WK    &  $871$ & $485$        &  $9,421$ & $976$     &    $32$ & $664$  &  $5,969$  & $350$ & $958$    & 
$\begin{array}{cc}
    >1 & 34\% \\
    >5 & 96\% \\
 \end{array}$\\\hline
\end{tabular}
\label{tb:stats}
\end{table*}

\section{Datasets}
\label{sec:datasets}
We have analyzed the discussion cascades of four websites.  In the following
paragraphs we give a more detailed description of the datasets and the
corresponding websites.  Global descriptive statistics can be found in
Table~\ref{tb:stats}.

%\vspace{.05cm} 
\textbf{Slashdot (SL)} : Slashdot\footnote{\url{http://slashdot.org/}}
is a popular technology-news website created in 1997 that publishes
frequently short news posts and allows its readers to comment on them.
Slashdot has a community based moderation system that awards a score
to every comment and upholds the quality of discussions. The comments
can be nested which allows us to extract the tree structure of the
discussion. A single news post triggers typically about 200 comments
(most of them in a few hours) during the approx. 2 weeks he is open
for discussion. Our dataset contains the entire amount of discussions
generated at Slashdot during a year (from August 2005 to August 2006).
See~\cite{gomez08} for more details about this dataset.
%about 2 million comments to 10,000 different news post.  

\textbf{Barrapunto (BP)} :
Barrapunto\footnote{\url{http://barrapunto.com/}} is a Spanish version of
Slashdot created in 1999. It runs the same open source software as Slashdot,
making the visual and functional appearance of the two sites very similar. They
differ in the language they use and the content of the news stories displayed,
which normally does not overlap. The volume of activity on Barrapunto is
significantly lower.  A news story on Barrapunto triggers on average around 50
comments. Our dataset contains the activity on Barrapunto during three years (from
January 2005 to December 2008).
% than the one of its %English counterpart
%, about $7,500$ posts which receive a total of approx. $400,000$ comments.

\textbf{Meneame (MN)} :
Meneame\footnote{\url{http://www.meneame.net/}} is the most successful
Spanish news aggregator. The website is based on the idea of promoting
user-submitted links to news (stories) according to user votes. It was
launched in December of 2005 as a Spanish equivalent to \verb|Digg|.
The entry page of Meneame consists of a sequence of stories recently
promoted to the front page, as well as a link to pages containing the
most popular, and newly submitted stories. Registered users can, among
other things: \emph{(a)} publish links to relevant news which are
retained in a queue until they collect a sufficient number of votes to
be promoted to the front page of Meneame,
%.  The promotion of links
%follows a relatively complex set of rules 
\emph{(b)} comment on links sent by other users (or themselves),
\emph{(c)} vote ({\it menear}) comments and links published by other users.
Contrary to both BP and SL, Meneame lacks an interface for nested comments.
Comments are displayed as a list so that the tree structure is hidden
%which are displayed as a list. 
%so that the tree structure is hidden to the user.  
However, the tag \verb|#n| can be used to
indicate a reply to the $n$-th comment in the comment list and
to extract the tree structures we analyze in this study.  To focus on the most
representative cascades, we filter out stories that were not promoted, that is
marked as discarded, abuse, etc.  Our dataset contains the promoted stories
and corresponding comments during the interval between 
Dec. 2005 and July 2009.

\textbf{Wikipedia (WK)} : The English
Wikipedia\footnote{\url{http://en.wikipedia.org}} is the largest
language version of Wikipedia. Every article in Wikipedia has its
corresponding \textit{article talk page} where users can discuss on
improving the article.  For our analysis we used a dump of the English
Wikipedia of March 2010 which contained data of about 3.2 million
articles, out of which about 870,000 articles had a corresponding
discussion page with at least one comment. In total these article
discussion pages contained about 9.4 million signed comments. Note
that the %AKht 1st note
comments are never deleted, so this number reflects the totality of
comments ever made about the articles in the dump. The oldest comments
date back to as early as 2001. Comments who are considered a reply to
a previous comment are indented, which allows to extract the tree
structure of the discussions. Note that Wikipedia discussion
pages %AKht 2nd note, remove one
contain, in addition to comments, structural elements such as
subpages, headlines, etc. which help to organize large discussions. We
eliminate all this elements and just concentrate our analysis on the
remaining pure discussion trees. More details about the dataset and
the corresponding data preparation and cleaning process can be found
in~\cite{Laniado2010}. %AK better add it again.
For our experiments we selected a random subset of $50,000$ articles
from the entire dataset. Results did not vary significantly when using
different random subsets of the data.

\subsection{Global analysis}

% SL: 93638  users,  0.9899 cascades where at least one user repeated.
%           max size 1031 users (1312, 1174)
% BP: 6864   users,  0.8476 cascades where at least one user repeated.
%           max size 180 users (1040, 345)
% MN: 53297, (53877 with post) users,  0.3393 for >2 and 0.6925 for >5
%           max size 1021 users (2718) 
% WK: 266493 users,  0.8115 for >2 and 0.9614 for >5
%           max size 3616 users (24081)
In this section we give a brief overview about some general characteristics of
the four datasets. Several indicators are shown in Table~\ref{tb:stats}. As
columns $4$ and $5$ show, the biggest observed discussions can be composed of
hundreds of comments and propagate across hundreds of users.  We find the
biggest discussion in Wikipedia, involving $5,969$ users and $32,664$
comments. In Barrapunto, however, the biggest conversation comprised $180$
users and $841$ comments.

%A global description of the datasets is shown in Table 1. %\ref{tb:stats} As As
%Column $4$ and $5$ show that the largest cascades can be composed of hundreds
%of nodes and propagate across hundreds of users too.  The largest cascade
%corresponds to the Wikipedia with $3,616$ users involved and $28,701$ nodes.
%The largest cascade of Barrapunto affects $180$ users only and has $1,040$
%nodes.

It is interesting to consider this quantity relative to the size of the
underlying social network (compare columns 5 and 6, where we indicate the total
number of users during the crawled period). We see a remarkable fact: the
percentage of users affected by the largest cascade is very small.  In
particular, it varies from a $1.1\%$ for Slashdot and $2.6\%$ in Barrapunto,
the dataset which we saw that presented the smallest cascade in absolute terms.
Globally, these results show that even the largest cascades only affect a very
small portion of the entire underlying social network.
\begin{figure}[!b]
\begin{center}
\includegraphics[angle=-90,width=.99\columnwidth]{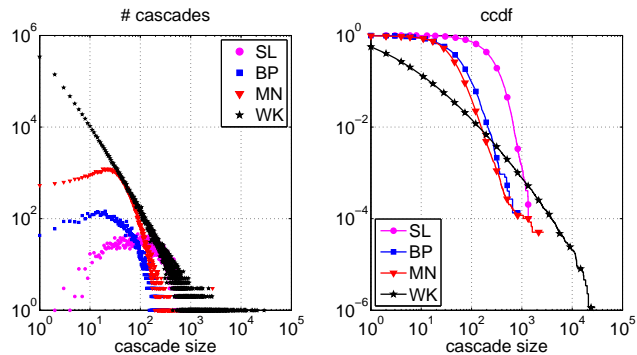}
\end{center}
\caption[Example]{Cascade sizes for the different datasets}
\label{fig:sizes}
\end{figure}

A characteristic feature of discussion cascades is the high frequency
of user participation. Evidence of this is provided in column $7$,
where we show the percentage of cascades in which at least one user is
involved more than once for cascades with more than two nodes (for MN
and WK, we also show the percentage for cascades with more than five
nodes).  With the exception of Meneame, all datasets show very high
values. In Slashdot, practically all posts contained at least one user
who commented more than once (considering only registered users). An
important consequence of this fact is that information diffusion may
not be properly explained using epidemic models such as SIR
(susceptible-infected-recovered) models unlike in other scenarios like
photo popularity \cite{flickr2} or fanning pages \cite{facebook}.
%% Extend
%\cite{twitter}
\begin{figure*}[!t]
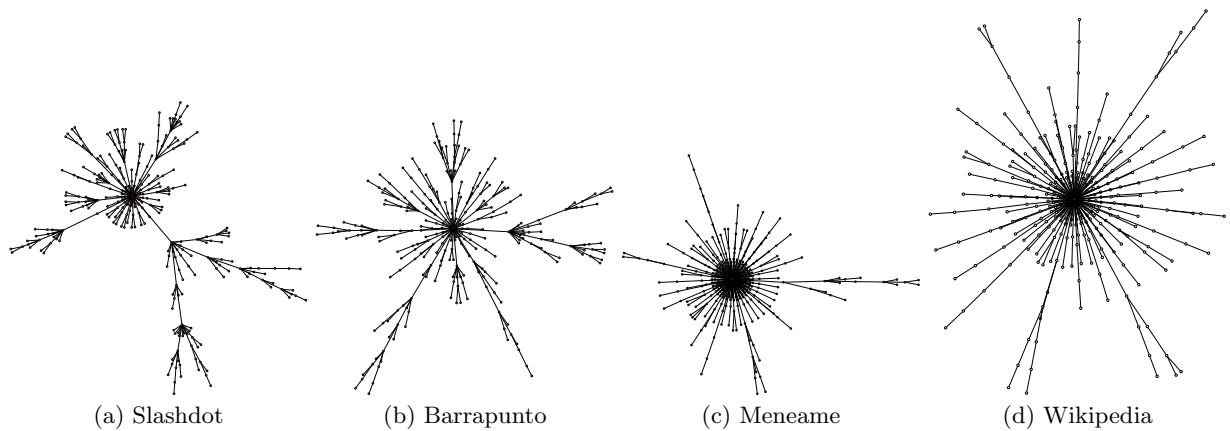

\begin{center}
 \subfigure[Slashdot]{\includegraphics[width=.47\columnwidth]{figures/sl_synt_tree.dot.eps}}
 \subfigure[Barrapunto]{\includegraphics[width=.47\columnwidth]{figures/bp_synt_tree.dot.eps}}
 \subfigure[Meneame]{\includegraphics[width=.47\columnwidth]{figures/mn_synt_tree.dot.eps}}
 \subfigure[Wikipedia]{\includegraphics[width=.47\columnwidth]{figures/wk_synt_tree.dot.eps}}
\end{center}
\caption[Example]{Examples of synthetic discussion cascades.}
%AKht maybe the figure description could be a bit more detailed.
\label{fig:synt}
\end{figure*}

Figure \ref{fig:sizes} shows the distribution of the cascade sizes of the four
datasets.  As expected, all distributions are positively skewed, showing a high
concentration of relatively short cascades and a long tail with large cascades.
However, although all distributions are heavy tailed, we clearly see a
different pattern between the three news aggregators and the Wikipedia.
Whereas SL, BP and MN present a distribution with a defined scale, the
distribution of cascade sizes of Wikipedia is closer to a scale-free
distribution, in line with the cascades found in weblogs \cite{Leskovek07} and
USENET~\cite{Kumar}.  We remark that, even in the Wikipedia case, the power-law
hypothesis for the tail of this distribution is not plausible via rigorous test
analysis: we obtain an exponent of $2.17$ at the cost of discarding $97\%$ of
the data.
%The presence of a scale is not always clear in the news websites.

We also observe a progressive deviation from websites with a well
defined scale such as Slashdot, which could be described using a
log-normal probability distribution, towards websites with less defined
scale such as Meneame, which may show a power-law behavior for cascade
sizes $>50$.  Barrapunto falls in the middle and, interestingly, is
more similar to Meneame than to Slashdot.

The previous considerations imply that, in general, a new post in Slashdot can
hardly stay unnoticed and will propagate almost surely over several users.
Conversely, most of the news in Meneame will only provoke a small reaction and
reach, if they do, a small group of users. Compared with Wikipedia, we can say
that Meneame is the news aggregator which has most similarities with it.

Figure \ref{fig:real} illustrates the different types of cascades which we
found. We plot representative cascades with similar sizes selected randomly
from each of the four datasets. For Slashdot we can see that the chain reaction
is located mainly on the initiator event (direct reactions), but some nodes also
have high degree, resulting in bursty disseminations. We could say that after a
news article is posted, the collective attention is constantly drifting from
the main post to some new comments % which are highly scored 
%AKht it is not necessary fro them to be highly scored, better omit
% this
which become more
popular.  In Barrapunto we observe similar structures, although their
persistence is less noticeable.  On the contrary, Meneame is characterized by
having high concentration of nodes at the first level together with rare but
long chains of thin threads.  
%This seems to represent a pattern where most
%people just leave a comment which often is not replied by other users, but
%which sporadically 
% AKht it might be better to say that 
This represents a pattern where only a few comments receive multiple replies,
but that sporadically can trigger a long dialog between a few users.  We note that
this phenomenon might be caused by the fact that the cascade tree
and, more importantly, the number of replies a comment receives are 
%AKht and more importantly 
hidden in the interface of Meneame.  Finally, the case of Wikipedia is very
similar to Meneame, but with even longer, more frequent and finer threads of
nodes with very low degree.

\section{Results}
\label{sec:results}
In this section we validate the proposed model by comparing the real cascades
to the ones generated using the model.

\subsection{Model validation description}
\label{sec:validation}
We use the cascades from the four datasets to validate the proposed PA model
with bias.  The parameters are optimized for each dataset independently using
the entire dataset and we generate the same number of synthetic cascades as the
number of real cascades extracted from each dataset. 
An alternative validation would be to use a train-test paradigm on each dataset
independently to prevent overfitting.  For simplicity, and since the goal of
this study is to characterize the different datasets instead of minimizing the
generalization error of new threads sampled from the model, we prefer to use the
entire datasets for learning.  \footnote{The estimated parameter values did not vary
significantly using different, sufficiently large random subsets of the data, as
the outcomes of a cross-validation (train-test) procedure would have produced.}
%AKht the last half-sentence is not clear}

The size of each synthetic cascade is pre-determined drawing a pseudo-random
number from the empirical distribution of cascade sizes (see Figure
\ref{fig:sizes}).  We calculate the following quantities from the empirical
data and from the synthetic cascades produced by the model:
%\vspace{-1mm} %AK added this
\begin{packed_description} %AK packed
    \item[Root node degree probability distribution:]
    Each cascade has a root degree, which is the number of \emph{direct}
    contributions to the root.
    \item[Total degree distribution:]
    We consider the degree probability distribution of any node, %AKht removed also
    without differentiating root versus non-root nodes.
    \item[Subtree sizes distribution:]
    For each non-root node, we compute the probability distribution
    of the total number of its descendants.
    \item[Mean node depth:]
    Each non-root node belongs to one level of the cascade. We compute the mean
    over all the levels of all the nodes.
  \item[Size - Proportion of direct reactions]: We compute the
    relation between the size of a cascade and the proportion of
    direct reactions to the root and analyze if they are correlated.
\end{packed_description}
%The distinction between root and non-root degree distributions is motivated
%by the fact that, in all datasets except Wikipedia, 
%we observed very different
%behaviors 
%of the first and the second measures 
%The choice of is not p
%Empirically, we observe different for all datasets except for the Wikipedia.
%
%Comment on the measures for non-root nodes.
%The choice of this two measures is justified because.
%All datasets, except Wikipedia show a clear distinction between the processes
%
%Therefore, are not just ad-hoc measures but 

\subsection{Structure of the cascades}

\begin{figure}[!t]
\begin{center}
\includegraphics[width=.85\columnwidth]{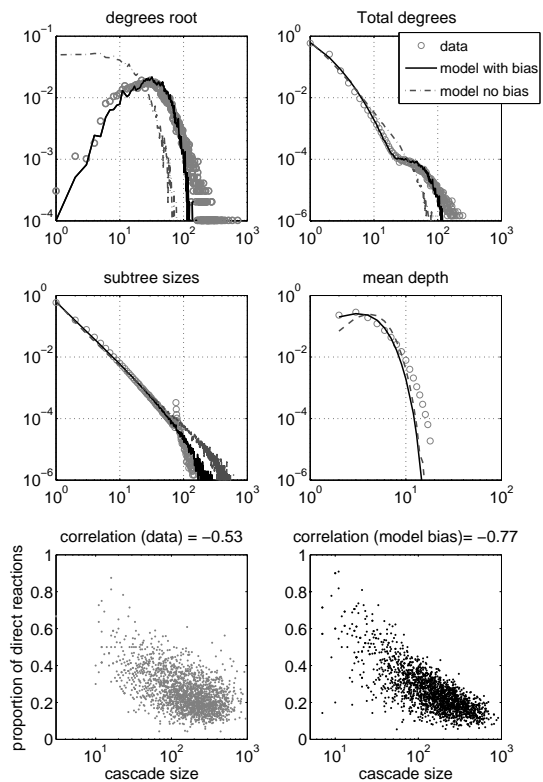}
\end{center}
\caption[Slashdot]{Model validation for Slashdot}
\label{fig:sl_fit}
\end{figure}
\begin{figure}[!t]
\begin{center}
\includegraphics[width=.85\columnwidth]{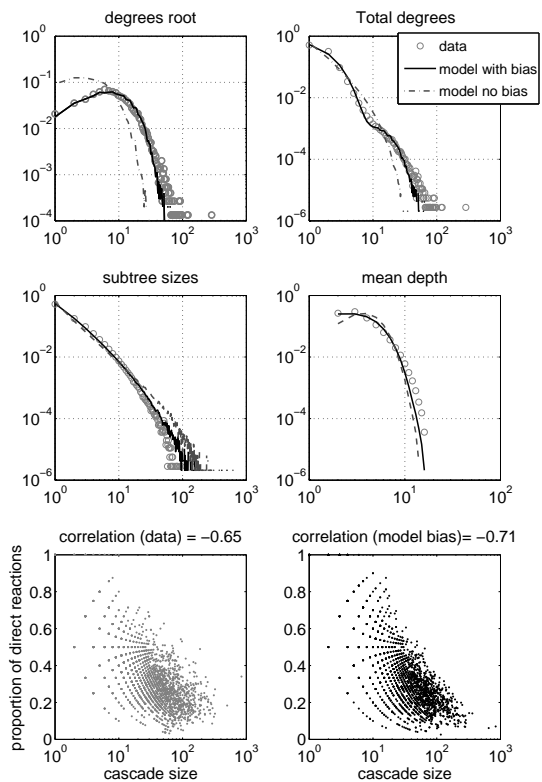}
\end{center}
\caption[Barrapunto]{Model validation for Barrapunto}
\label{fig:bp_fit}
\end{figure}
\begin{figure}[!t]
\begin{center}
\includegraphics[width=.9\columnwidth]{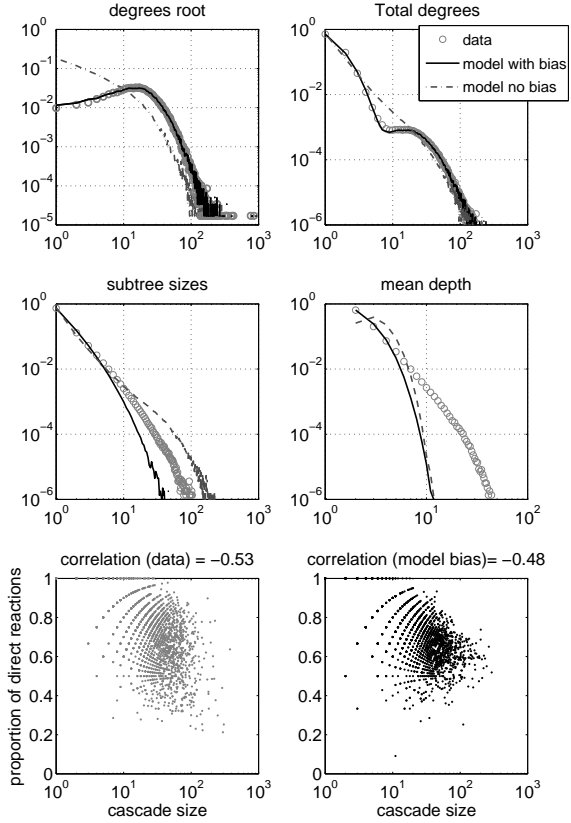}
\end{center}
\caption[Meneame]{Model validation for Meneame}
\label{fig:mn_fit}
\end{figure}

\begin{figure}[!t]
\begin{center}
\includegraphics[width=.9\columnwidth]{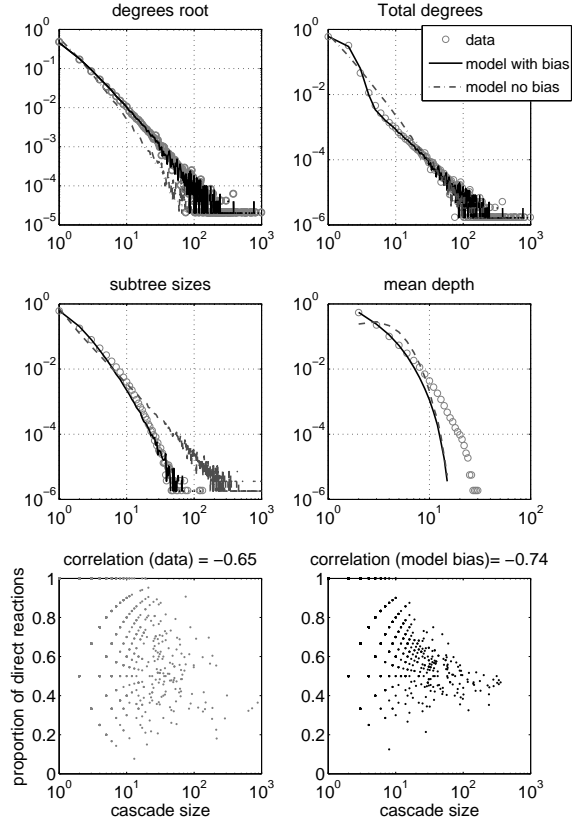}
\end{center}
\caption[Wikipedia]{Model validation for Wikipedia}
\label{fig:wk_fit}
\end{figure}
%First, we see that the model with bias is able to capture the degree
%distributions of the root nodes, even though each dataset shows a different
%profile for this quantity.  The simple model without bias generates a very
%different distribution, where root nodes with higher degree have always higher
%probability.  Parameter $\beta$ is determinant to predict accurately this
%distribution.

Figures \ref{fig:sl_fit}, \ref{fig:bp_fit}, \ref{fig:mn_fit} and
\ref{fig:wk_fit} show plots of the previous quantities for each dataset and the
outcomes of both PA models with and without bias to the root.

%\begin{enumerate}
%\item \textbf{Assumption A1} Puta barata \label{puta}
%\item \textbf{Asuptio A2} Puta casa
%\end{enumerate}

Overall, the model with bias is able to capture reasonably well all the
measured properties, except the mean depth.  In particular the degree
distributions of the root nodes are very accurately reproduced, even though
each dataset exhibits a different profile (see top-left plot of the figures).
For this quantity, the difference between using or not a bias term is clearly
manifested.  A model without bias systematically produces degree distributions
too skewed for the non-root nodes and with too short tails for the root nodes,
and is not able to capture qualitatively the shape of the total degree
distribution (see top-right plots of the figures).
%Due to, we can observe that
%the tail of the degree distributions , since is Once this effect is removed, the observed
%degrees do not reach more than a few decades.

A similar behavior is observed in the correlations between the log-size of the
cascade and the proportion of direct reactions (bottom plots of the figures).
Although the scatter plots differ substantially across datasets, the model with
bias is able to reproduce them qualitatively, which is not the case for the
model without bias (data not shown).

% Introducing a bias term favoring the root node is fundamental to match..
% parameter $\beta$ is fundamental to reproduce this distribution,
The model with bias also generates correct subtree sizes in general, with the
exception of Meneame, which we postulate is caused by the particularities of
the platform (see Section~\ref{sec:discussion} for details).  On the contrary,
the model without bias systematically produces longer tails than the real ones.
%We also see a slight deviation on the tails of the non-root degree
%distributions, more noticeable in Meneame and the Wikipedia, for which the
%distributions are marginally less skewed.  Nevertheless, the model with bias
%produces much better approximations than the model without bias, which again
%systematically produces longer tails than the real ones.  

Both models tend to produce shorter tails for the mean depth distribution in
all datasets. This seems to be a current limitation of the model.  Although for
Slashdot and Barrapunto this deviation is not very severe, for the other two
datasets we observe clear discrepancies at the tail of the distributions.
Notice that in this case, the model without bias is unable even to reproduce
the probability mass corresponding to the first values of the distribution.
We will return to this point in Section~\ref{sec:discussion}.

%The proportion of direct reactions (bottom plots) is .
%The outcomes of the model without bias (data not shown).

%		Meneame deserves a special attention. As figure \ref{fig:} suggests,
%		subtree sizes a
%		we believe that again this is caused by the fact that the software
%		platform strongly determines.  We also compare the between  the size of the
%		cascade and the proportion of nodes that link directly to the root.  Meneame 

To conclude this section, we show in Figure \ref{fig:synt} the synthetic
counterpart of Figure \ref{fig:real}, where we plot representative cascades
with similar sizes selected randomly from each of the four synthetic datasets.
We can see that the generated cascades present a strong
resemblance with the real ones.

%   However, the degree distributions
%   of Slashdot are 
%   %The degree distributions of the non-root nodes are always less skewed than the
%   %previous ones and resemble a stretched exponential.  
%   %observe a slight deviation at the tails. In the case of Barrapunto we observe a
%   %pattern of alternation which could be produced by a minor artifact in the data.
%   Both models robustly overcome this problem and generate degree distributions
%   %which explain the observations satisfactorily.  
%
%The subtree sizes distributions are also properly approximated with the exception
%of Meneame, again 

%\begin{enumerate}
%\item subtree sizes: most similar across all the datasets.
%Model without bias produced heavier tails in all datasets specially in Meneame
%and wikipedia, alpha>1.
%\item The mean depth distribution is systematically underestimated.
%Although for Slashdot and Barrapunto is not that severe, for the other two datasets,
%we observe a clear different behavior in the tail.
%This seems to be the limitation of our model.
%\item Meneame shows limitations of the model, which is unable to capture the
%proper subtree sizes and the mean depths. 
%\item heterogeneity in the correlations. Well captured again with the model with bias.
%\end{enumerate}

%   plots fits: d
%   as if we compare 
%   The model without bias produces very different distributions.

\subsection{Evolution of the cascades}
After having compared the main structural properties of the synthetic trees
with the real ones, we now investigate whether the
PA model with bias is also able to reproduce the growth process of the
cascades.  In other words, if we take intermediate snapshots of the cascades
during their evolution, how close match the synthetic trees their archetypes?

To this end we record two quantities: the \textbf{width} (maximum over the
number of nodes per level) and the \textbf{mean depth} of the trees every time
a new node is added (at every timestep).  Note that the timesteps in the model
do not coincide with the actual time differences between the comments. They
just reflect the sequence of the comments attaching to the cascade.  In
reality, information spreading is conditioned to the large heterogeneity
present in human activity, for instance induced by circadian cycles, which
results in information transmission speeds governed by subexponential
distributions, i.e.  log-normals or power-laws \cite{iribarren,
kaltenbrunner_LAWEB2007, malmgren}.  Capturing the growth process of the real
cascades is therefore a challenging task for our model.

%(for statistics about the actual temporal growth of the cascades
%in the Slashdot dataset see~\cite{kaltenbrunner_LAWEB2007}). 

\begin{figure}[!t]
\begin{center}
\includegraphics[angle=-90,width=.69\columnwidth]{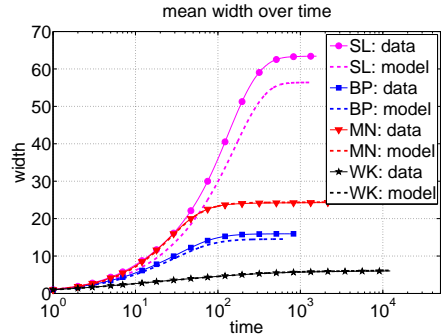}
\end{center}
\caption[width evolution]{Evolution of the width}
\label{fig:width_time}
\end{figure}

\begin{figure}[!t]
\begin{center}
\includegraphics[angle=-90,width=.66\columnwidth]{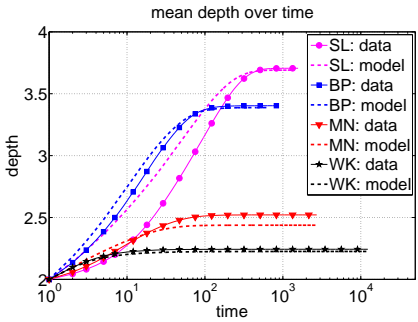}
\end{center}
\caption[mean depth evolution]{Evolution of the mean depth}
\label{fig:mean_depth_time}
\end{figure}

The average overall width and depth evolution curves are presented in
Figures~\ref{fig:width_time} and~\ref{fig:mean_depth_time} for all datasets
comparing the original cascades (continuous lines with symbols) with the
biased model (dashed lines). We observe a nearly perfect coincidence between
our model and the data in the evolution of the width of the discussions
(Figure~\ref{fig:width_time}), for three of the four datasets. Only in the case
of Slashdot the model underestimates the width of the tree, although it still
reproduces the same curve shape if normalized by the final depth.

The picture in the case of the mean depth (Figure~\ref{fig:mean_depth_time}) 
is less favorable, but still shows a reasonable coincidence of our
model with the data. In the case of Wikipedia, although the model 
underestimates the mean depth, it reproduces a rescaled version of it.  The
other datasets show a similar profile. Initially, the synthetic trees 
are too deep and the mean depth is overestimated. This deviation is
corrected at some point and then the opposite effect takes place: when
the depth of the synthetic trees saturates, the depth of the real ones still
grows.  The initial deviation is specially severe in Slashdot, for which
remarkably the final mean depth is very close to the one of the real
%AKht remarkably or eventually, but not both
cascades.  A possible way to overcome this problem is discussed in Section~\ref{sec:discussion}.
%       Again the worst coincidence between our model and the data
%       is found for Slashdot, where the model systematically overestimates the
%       depth of the discussions in the beginning but finally does not reach
%       the final average mean depth over all cascades. The reason for this
%       different behavior in the case of Slashdot may be found in 
%       % AK: Add something here
%   The remaining two datasets follow initially quite closely the depth
%   evolution curve but do not reach the final maximum depths
%   found in the data. 

\subsection{Interpretation of parameters}
\label{sec:params}
\begin{figure}[!b]
\begin{center}
%\subfigure[\label{fig:params}Four different datasets]{\includegraphics[angle=-90,width=.49\columnwidth]{figures/parameters3.eps}}
\subfigure[\label{fig:params}Four different datasets]{\includegraphics[angle=-90,width=.49\columnwidth]{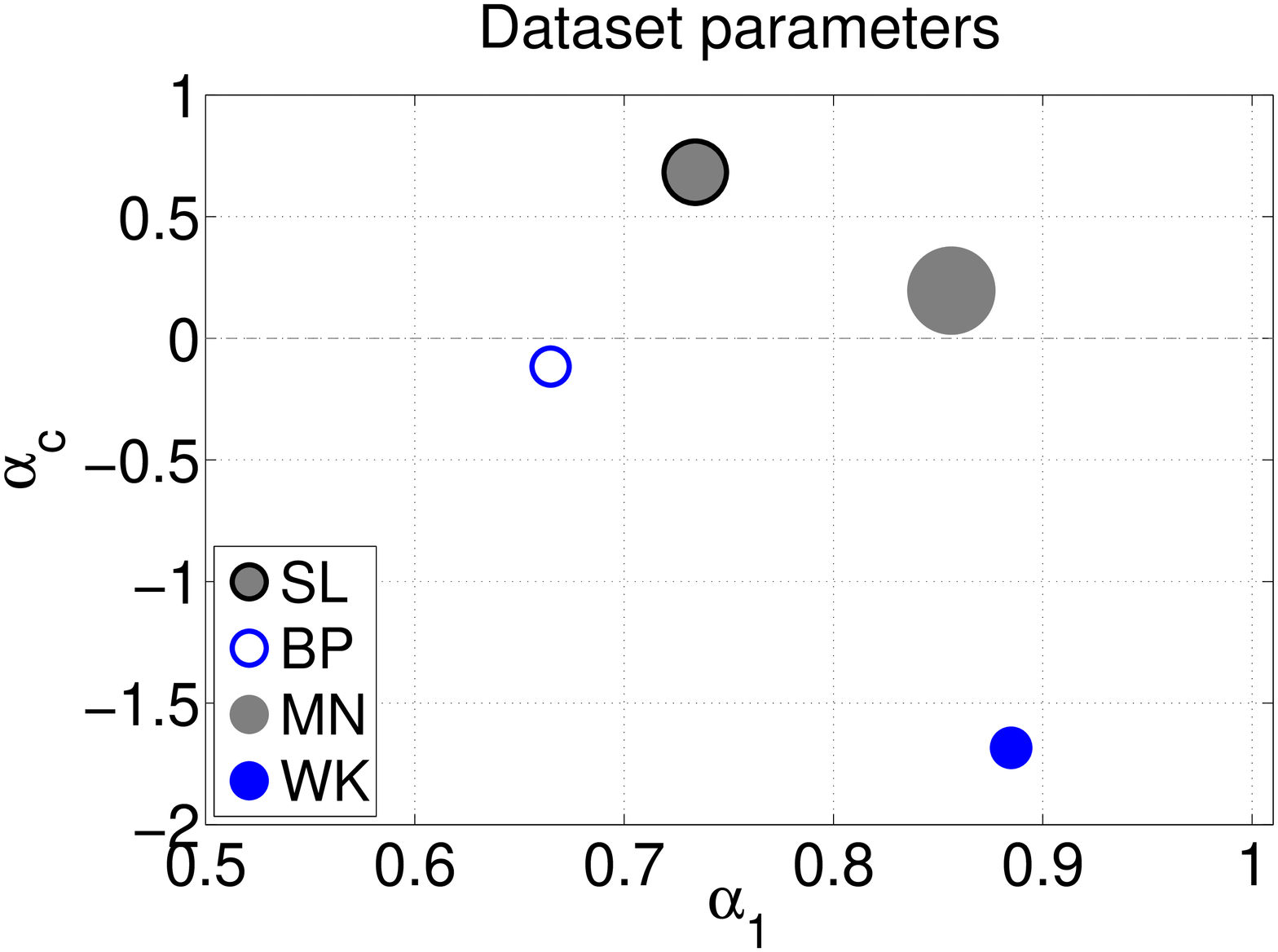}}
\subfigure[\label{fig:SDparams}Slashdot topics]{\includegraphics[angle=-90,width=.49\columnwidth]{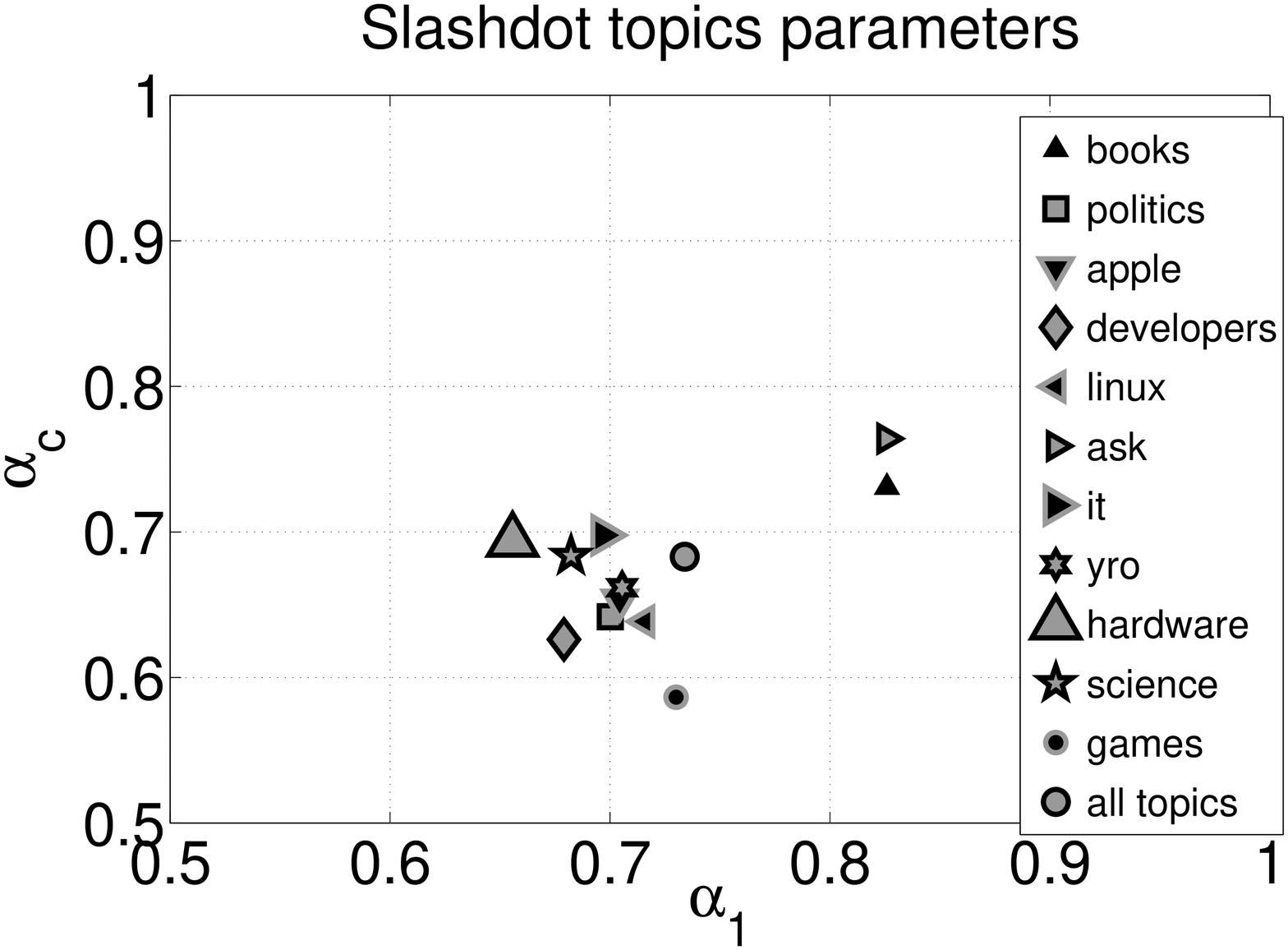}}
\end{center}
\caption[Example]{Comparison of parameter values for (a) the different
  datasets and (b) the topics of Slashdot. Marker sizes encode $\beta$ differently in (a) and (b).}
%\label{fig:params}
\end{figure}

Can we derive conclusions about the communication habits which characterize each
website based on the obtained parameters which best fit each model?  Figure
\ref{fig:params} shows the optimal parameter values for each dataset in a three
dimensional plot, where the horizontal and vertical axis correspond to
$\alpha_1$ and $\alpha_c$ respectively and the size of the marker to the value
$\beta$. Table~\ref{tb:params} shows the same values numerically.

The role of the exponents $\alpha_1$ and $\alpha_c$ in the model is to quantify
the degree of preferential attachment of the root node and the non-root nodes
respectively.  The higher their values, the more relevant is the 
popularity to determine the attractiveness to new nodes in the cascade.
For instance, values very close to zero imply a random cascade where new nodes
are linked to existing ones with uniform probability.  
We can use the established theoretical results described in Section \ref{sec:model}
to characterize the websites under study.  
%It is well known that in that case we recover a random
%network where the degrees of the cascade nodes are exponentially distributed.
%On the contrary, higher exponent values indicate that the popularity of a node
%is determinant to characterize the growth process. In our case, an exponent
%equal to one will result in power law distributed degrees.

The different exponents $\alpha_1$ are all sublinear ($<1$),
relatively high, and very similar in all datasets, indicating a strong
PA in the root process of all cascades. The two lowest values for this
quantity are observed for Barrapunto and Slashdot.
%, in agreement with the known similarities of both websites.  
On the other hand, Meneame and Wikipedia present a higher and almost
identical value, suggesting a very similar role of the root nodes in
the PA mechanism of both websites.
\begin{table}[!tb] \centering
\caption{Optimal parameters}
\begin{tabular}{|c|r@{.}l|r@{.}l|r@{.}l|}
\hline 
    dataset &
    \multicolumn{2}{|c|}{$\alpha_1$} &
    \multicolumn{2}{|c|}{$\alpha_c$} &
    \multicolumn{2}{|c|}{$\beta$} 
\\\hline
SL    &   $0$ & $734$       &   $0$  & $683$   &  $1$ & $302$ \\\hline
%BP    &   $0$ & $678$       &   $0$  & $400$   &  $0$ & $950$ \\\hline
BP    &   $0$ & $665$       &   $-0$  & $116$   &  $0$ & $781$ \\\hline
MN    &   $0$ & $856$       &   $0$  & $196$   &  $1$ & $588$ \\\hline
WK    &   $0$ & $884$       &   $-1$ & $684$   &  $0$ & $794$ \\\hline
\end{tabular}
\label{tb:params}
\end{table}

A clear segregation between the group of three news media websites and the
Wikipedia is manifested on Figure~\ref{fig:params} in the value of $\alpha_c$.
%The first group, in order of increasing value of $\alpha_c$, consists of
%Barrapunto, Meneame and Slashdot. 
Slashdot, has the highest value, $\alpha_c \approx 0.68$, even higher
than $\alpha_1$ for Barrapunto.  It is also very similar to $\alpha_1$
for the same dataset.  This similarity may capture the special quality
of the Slashdot comments.  In a sense, good comments may behave like
posts and may act eventually as effective initiator of information
diffusion cascades.

The smaller value of $\approx 0.2$ found for Meneame indicates that
the diffusion of the news comments in this website is closer to a random
process. This can be again influenced by the lack of explicit
information about the popularity of a given comment. The same is true
for Barrapunto although its value of $\alpha_c\approx -0.1$ is slightly
negative, indicating a slight \emph{inverse} PA process.

%Which however is much more 
However, such an \emph{inverse} PA process is much more prominent in
the case of Wikipedia: whereas its $\alpha_1$ is high, indicating a
strong PA in the root process in agreement with the other datasets,
its $\alpha_c$ is negative and has the largest absolute value of all
four datasets.  What are the implications of this result?
% An interpretation is that the process of cascade growth in Wikipedia
% is  actually an  \emph{inverse} PA  process.  The  suggested process
% here is that
Once a comment on a Wikipedia article has been originated, it will
derive in a collaborative reciprocal chain between a very reduced
group of contributors. So once a node has received a reply it will be
three (exactly: $2^{1.686}=3.2$) times less likely to receive
another one than the replying new comment itself.  In other words,
nodes with degree equal to one (leaf nodes) are much more likely to be
linked with a new node than nodes which higher degrees.

%In this case, a new contribution which has been
%already decided not to be linked directly to the root, will be more likely
%attached to a node with less links.

%causing an update in the evolving
%article, for instance.  The new node will replace the previous one and will be
%more likely to be linked with subsequent contributions.

Finally, the parameter $\beta$ acts as a weight which expresses
the trend towards the root of the cascade in relation to the subsequent nodes.
It is especially important in the beginning of the cascade, when the degree of the
root node is low, and determines whether initially many nodes attach to the root or
rather to one of the first comments. We observe thus that Meneame shows the
largest initial predominance of direct reactions, while Wikipedia gives higher
probability mass to the comments, allowing thus large chains already with a
small number of nodes. The values for Slashdot and Barrapunto lie in between
indicating an intermediate initial preference for the root node, showing
Barrapunto a higher probability for early reply-chains than Slashdot.
%In this case, Wikipedia has the smallest weight, and Meneame and Slashdot
%present the highest values.
%The parameter $\beta$

% AK we should add something more the difference between SL and BP,
% e.g.: In SL the preference for the root node is encoded on beta rather
% than in alpha.  th

% Slashdot and Meneame appear close to each other than are the ones
% with higher weights, and Barrapunto and Wikipedia.higher values of
% $\beta$ correspond to websites for which the which are driven mostly
% are signatures of websites which rely
%\begin{figure}[!b]
%\begin{center}
%\includegraphics[angle=-90,width=.65\columnwidth]{figures/parameters_topics_sl2.eps}
%\end{center}
%\caption[Example]{Parameters obtained for the different datasets
%for initial attractiveness $c=1$.}
%\label{fig:params}
%%AKht same label as figure 11.
%\end{figure}

What would be the scenario if one tries to explain the cascades using
the simple model without bias to the root?  We also fit the simple PA
model to the data. In that case, we would infer mistakenly that both
Meneame and Wikipedia systems are in the ``condensation''
regime~\cite{kaprivsky}, since their exponents ($1.360$ and $1.161$
respectively) are larger than one.
%AKht where do these values come from, the are not reported anywhere else.
%Vicen: Added "We also fit the simple PA model to the          |  simple
%For Barrapunto and Slashdot would also be higher ($0.820$ and $0.975$ respectively).
%are still
%sublinear, the values for Meneame ($1.360$) and Wikipedia ($1.161$) are both larger than
%one, and we would infer mistakenly 
%In the latter models, one node of the cascade (not necessarily the post) acts
%as the "gel" node and attracts many comments.  However, this mechanism, as we
%have established in the previous section, does not provide valid explanation
%for the observed cascades.

% Summarizing, the optimal parameters allow an
% interpretation of the communication habits of each social space.  This
% representation also leads to different classification as a function of the
% parameters. For instance, $\alpha_1$ separates Slashdot and Barrapunto from
% Wikipedia and Meneame, and $\alpha_c$ splits up the Wikipedia from the three
% news media.

The next question we want to answer is how stable are these parameters
within the same site, i.e. if we split for example different
discussions according to their category, do we obtain similar
parameters? We investigate this question for the topic categories in
Slashdot (we only consider categories with more than 100 discussions)
and find (see Figure~\ref{fig:SDparams}) that the majority of
topics has a set of parameters which is close to the one obtained
for the entire website. 

This is remarkable given the heterogeneous picture that is observed if the
depths and widths of the discussions are
considered~\cite{GonzalezBailonJIT2010}. So it seems that, although the amount
of comments which attracts the different categories may be different, the
actual structure of the discussions follows a very similar pattern. However, 
if we consider the $\alpha$ parameters, we observe three outliers from
the main cluster. The topics ``books'' and
``ask'' have much larger values,
indicating a more experienced preferential attachment behavior, while the
topic ``games'', on the other hand, has the largest difference between
$\alpha_1$ and $\alpha_c$.  It seems that in this topic category direct comments to
the root node are more frequent\footnote{This category is also the one with the
shallowest trees~\cite{GonzalezBailonJIT2010}.} while in the two other outliers
also comments seem to be able to attract a reasonable amount of replies. It is
also interesting to observe the differences in the values of $\beta$, where we find the largest
trend toward the root for ``hardware''  and the smallest for
``books'' articles.
%, the two categories with the largest
%and lowest initial preference for the root.
%In conclusion we can find some differences between the categories of a
%site, but the differences are small compared to 

Summarizing, the optimal parameters permit an interpretation of the
communication habits of each social space and are relatively stable across
different categories within a site. This representation also leads to different
classification as a function of the parameters.  The bias to the root 
node is crucial to separate Slashdot and Barrapunto from Wikipedia and Meneame
according to $\alpha_1$, and Wikipedia from the three news aggregators
according to $\alpha_c$.

%Related to this we would like to emphasize one final aspect. If we consider the
%model \emph{without} bias to the post where $\alpha_c=\alpha_1$ and $\beta=0$,
%the maximum likelihood values obtained for $\alpha>1$, where .
%However, this hypothesis is false, since the fits of , as the figures
%Once we decouple both global and local processes, all the become sublinear.

\section{Related work}
\label{sec:related}
Due to the increasing availability of empirical data on cascades, extensive
work is appearing with focus on how information cascades are propagated in a
social network.
%motivated from epidemiology.  information diffusion using disease spreading models.

At a statistical description level, information cascades have been analyzed in
detail for particular social spaces. Twitter cascades \cite{twitter} are
predominantly shallow and wide (maximum depth is 11).  Flickr \cite{flickr}
shows the remarkable phenomenon that popular photos spread slowly and not
widely. This is in harmony with our findings which report that even the largest
realizations reach a very small proportion of the social network.  
%The forum threads of USENET, which
%conceptually resembles most similarities with the discussion cascades of this
%work, have been analyzed in \cite{Mcglohon2}, where a measure is proposed which
%assigns ownership of posts to groups allows to infer interesting patterns about
%how content is diffused between groups.
%AKht I think this has nothing to do with our work we do not need to cite it.

%Our model should in principle be valid for such cascades.
%The Facebook Page diffusion has been
%analyzed in \cite{facebook}.  Although the structures analyzed in that study
%are not trees, one could extract cascades such as the ones modeled here.  In
%terms of the model proposed here, initiators could be modeled as direct
%reactions of a common root node.

%using a predictive model.
%All these studies show strong heterogeneities between 

% for instance, Slashdot \cite{gomez08},
Blog cascades have been analyzed in \cite{Leskovek07}.  Interestingly, although
one would expect blog cascades to share more similarities with the discussion
cascades existing in Slashdot or Meneame, it is the Wikipedia dataset which
shows most similar patterns to the blog cascades (see Figure \ref{fig:sizes}).
In \cite{blogs}, a model of both blogger (user) and cascades was presented
which reproduces global temporal and structural aspects of the blogosphere. We
note that the motivation of our work is rather different.  Whereas
\cite{Leskovek07,blogs} aims for finding the simplest, parameter-free model
able to describe both user network and cascade behavior, we look for a
parameterized model from which we can describe communication habits which
characterize a particular website (see Section \ref{sec:params}).  In contrast
to the blog data, the datasets considered here contain complete information of
the cascade evolution. In this sense, our data avoids selection bias which
strongly influences the estimation of these processes \cite{golub}.  In
\cite{golub}, a simple branching process (Galton-Watson process) is proposed
for modeling chain-letter cascades.  Although such a model may explain certain
characteristics such as depths distributions (after proper correction for
selection bias) it cannot capture the cascade evolution and assumes that all
degree distributions are independent, so its utility for our purposes remains
very limited.

%The most important difference between our approach and the aforementioned ones
%is that in our datasets a new comment can in principle choose its parent
%between the entire set of comments.  Further, (with the exception of Meneame)
During the development of this manuscript we learned of the work of
Kumar et al.~\cite{Kumar} which also presents a model for discussion
trees, called T-MODEL. The same study also considers other aspects of
the cascades such as the identities of each member of the
conversation.  Our work is focused on the cascade model, its parameter
estimation and validation on the four datasets.  The same or a
different authorship model as the one of~\cite{Kumar} could also be
built on top of the model proposed here.

%overlaps some of our results for three
%different datasets. % The main difference is that, instead of the
% probability model of Equation \eqref{eq:model}, their growth cascade
% model (T-MODEL) uses other three parameters ($\gamma,\tau$ and
% $\delta$) and evolves according to:
% \begin{align}
% \label{eq:tmodel}
% p(\pi_{t+1} = k| \boldsymbol{\pi}_{(1:t)})
% &= \frac{1}{Z_t}\left(\gamma d_{k,t}+\tau^{t+1-k}\right),\notag\\
% Z_t &
%     = \delta + \gamma(t-1)+\frac{\tau(\tau^t-1)}{(\tau-1)}.
% \end{align}
% % = \delta+\sum_{l=1}^{t}{\alpha d_{l,t}+\tau^{t+1-l}}
% or finishes with probability $\delta/Z_t$, Parameter $\gamma$ (named
% $\alpha$ in their paper) captures the popularity of a node, parameter
% $\tau$ the novelty and $\delta$ the size of the cascade. 
The T-MODEL is based on \emph{linear} preferential attachment only, and unlike
ours does not distinguish between root and subsequent nodes. However, it
includes a \emph{recency} term which allows it to capture qualitatively the
relation between the sizes and the depths of the cascades. 
%A further
%difference between the T-MODEL and our study is that it also includes a
%parameter for the termination of the discussion.
%[We believe that the size of a
%discussion should be treated independently of its structure as it will be
%heavily influenced by exogenous factors such as attention spans or the
%existence of other discussions competing for audience. Therefore we conditioned
%the cascade size to be distributed according to the data and]
  % We believe that the size of a discussion should be
  % treated independently of its structure as it will be heavily
  % influenced by external factors such as attention spans or the
  % existence of other discussions competing for audience. Therefore we
  % did not include such a parameter in our model.}.
% AK maybe we should add a citation to a model for this (something
% with blog-posts maybe, Götz, Leskovec, et al. ?)
Preliminary experiments indicate similar ability of our model in this
aspect.  Additionally, the bias to the root considered here clearly
permits to capture other quantities with higher accuracy, such as the
degree distributions or the subtree sizes.  This suggests that at
least for the datasets analyzed here our model performs better.  The
maximum likelihood estimation scheme presented here finds the best
parameters of a model given the data, and therefore allows to quantify
objectively the predictive power of different models.  Such a
comparison between the two cascade models and possible hybrid forms is
left for future research.

A further difference between the T-MODEL and our study is that it also
includes a parameter for the termination of the discussion. The
resulting termination probability of a discussion is independent of
its actual structure and could be substituted by any other model
encoding the popularity of discussions (e.g.~\cite{wu2007novelty}),
which could also be combined with our model.

\section{Discussion}
\label{sec:discussion}
% PAJA
We have presented a thoughtful analysis and comparison of the structure and
evolution of the different discussion cascades of three popular news media
websites and the English Wikipedia.  Our analysis highlights the
heterogeneities between the discussion cascades, which can be conditioned from
two factors, namely, the page design, or platform, and the audience.  Despite
this, we have given evidence that a simple model can capture most of the
structural properties and the evolution profiles of the real cascades with the
particularities of each dataset.  Further, we have derived a rigorous maximum
likelihood approach which considers the entire evolution of the cascade.  The
learned parameters of the model proposed here allow for a figurative
description that characterizes the communication habits of a website.

% We have shown evidence that Meneame
%	
%limitation appears from the inability of the model to
%reproduce the situation where older nodes gradually become less attractive than
%newer ones. This 
For some datasets, the model tends to produce too shallow cascades.  We
postulate that this occurs especially in mature discussions, where interaction
at the leaves only happens between a few individuals who start to reply
mutually to each other and increase the mean cascade depth considerably.  
A possible extension which could correct for that effect is focus 
of current research.

\subsection{Acknowledgments}
We wish to thank David Laniado and Riccardo Tasso for providing the
preprocessed Wikipedia dataset and \url{Meneame.net} for allowing to access an
anonymised dump of their database.  We also thank Mohammad Gheshlaghi and
Alberto Llera for useful discussions.
%\balancecolumns
\appendix

\section*{Log-likelihood function}
In this appendix we describe some considerations related to the log likelihood
function \eqref{eq:loglikZ} we want to minimize. Briefly, we show that the PA
model can be formulated as a probability distribution which belongs to the
exponential family.  Consequently, the optimization problem is convex, e.g.
has the convenient property that any local minimum is global. 

Without loss of generality we can assume that parameter $\beta_k$
is of the following form:
\begin{align*}
\beta_k^{\alpha_k} & := \exp\left(\beta'_k\right).
\end{align*}
%d_{k,t}^{\ak}      & := \exp\left(\ak\log d_{k,t} \right).
We can rewrite the PA model defined in Equation \eqref{eq:model} as:
%\begin{align}\label{eq:exp}
\begin{align*}
p(\pi_{t}=k|\boldsymbol{\pi}_{(1:t-1)}) &= 
\frac{1}{Z'_t}\exp\left(\beta'_k + \ak\log d_{k,t}\right),
\end{align*}
where $Z'_t = \sum_{l=1}^t{\exp\left(\beta'_l + \al\log d_{l,t}\right)}$.  This
probability distribution is equivalent to that of the Equation
\eqref{eq:model}, but expressed in terms of the exponential family.  The
log-likelihood function \eqref{eq:lik} can be rewritten as:
%\begin{align}\label{eq:loglikZ_exp}
\begin{align*}
\log\mathcal{L}'(\boldsymbol{\Pi}|\boldsymbol{\theta})&=
\sum_{i=1}^{N} \sum_{t=2}^{|\boldsymbol{\pi}_i|} 
\beta'_x+\alpha_x\log d_{x,t,i}-\log Z'_{t,i}(\boldsymbol{\pi}_i|\boldsymbol{\theta}),
\end{align*}
where
$Z'_{t,i}(\boldsymbol{\pi}_i|\boldsymbol{\theta})= \sum_{l=1}^t{\exp\left(\beta'_l + \al\log d_{l,t,i}\right)}$.
%$$\sum_{l=1}^{t}{\exp\left(\beta'_l \al\log d_{l,t,i}\right)}$$
%Since \eqref{eq:exp} belongs to the exponential family,
%the minimization of this function is a convex problem.
The Hessian of this function (matrix of second order partial derivatives) is
always positive semi-definite.

The presented method can therefore be applied to any set of observations which
can be expressed as a collection of parent vectors $\boldsymbol{\Pi}$ from
which the degrees of each node at each time-step can be obtained. 
%For
%instance, we can start with an initial $\theta_{\text{bias}}^0$, and update it
%in the direction of \eqref{eq:da1}, \eqref{eq:dac}, \eqref{eq:db} until we
%converge to the global optimum
%$|\theta_{\text{bias}}^{j+1}-\theta_{\text{bias}}^{j}| < \gamma$, for a small
%threshold $\gamma$. 
Once the minimization is performed, we can recover the original parameter
%$\beta_k$ using: $$\bk= \exp\left(\frac{\beta'_k}{\alpha_k}\right).$$ The basic
$\beta_k$ with $\bk= \exp\left(\beta'_k/\alpha_k\right).$ The basic
PA model is the special case where $\alpha =\alpha_1 = \alpha_c$.
%and the only
%update equation corresponds to \eqref{eq:dac} where the sum in the numerator
%starts at $l=1$.  

Note that the bias to the root node can be introduced:
\begin{packed_description} %AK packed
\item[(A)] Using two alphas $\alpha_1$, $\alpha_c$ but no $\beta$ ($\beta=0$).
\item[(B)] Using one alpha $\alpha = \alpha_1 =\alpha_c$ and $\beta$.
\item[(C)] Using two alphas $\alpha = \alpha_1 =\alpha_c$ and $\beta$ (the approach
presented in this manuscript).
\end{packed_description}
As expected, since model \textbf{(C)} uses more parameters than \textbf{(A)}
and \textbf{(B)}, the resulting likelihoods and fits are better. In particular,
the impact of adding $\beta$ as a parameter is notable in the approximated
measures related to the root node, for instance the root degree distributions.

Notice that the convexity does not imply uniqueness of optimal
parameter values.  It could happen that the same minimum is attained
for a large range of parameter
values. %, as we experienced with the T-MODEL.
We used as an optimization procedure the Nelder-Mead simplex algorithm
(implemented as \url{fminsearch} in Matlab) which is an unconstrained
non-linear direct search method that does not use numerical or
analytic gradients.  Starting from many different random initial
conditions, we did not find multiple optimal values in any of
the datasets, so we can conclude that the presented values for each
dataset are unique.

\bibliographystyle{plain}
%\bibliography{thesis}

%Generated by bibtex from your ~.bib file.  Run latex,
%then bibtex, then latex twice (to resolve references)
%to create the ~.bbl file.  Insert that ~.bbl file into
%the .tex source file and comment out
%the command \texttt{{\char'134}thebibliography}.
%%
%% This next section command marks the start of
%% Appendix B, and does not continue the present hierarchy
%\section{More Help for the Hardy}
%The www2010-submission.cls file itself is chock-full of succinct
%and helpful comments.  If you consider yourself a moderately
%experienced to expert user of \LaTeX, you may find reading
%it useful but please remember not to change it.
\balancecolumns % GM July 2000
% That's all folks!
\end{document}